\documentstyle[12pt]{article}
\begin{document}
\title{\large \bf On the underlying theory approach for quantum 
theories}
\author{Jifeng Yang$^*$ \\
 School of Management \\
Fudan University, Shanghai, 200433, P. R. China}
\footnotetext{$^*$ E-mail:jfyang@fudan.edu.cn}
\date{}
\maketitle
\begin{abstract}
Some explanations and implications of the underlying theory 
approach for quantum theories (QM or QFT) are discussed and
suggested. This simple idea seems to have significantly
nontrivial effects for our understanding of the quantum theories.
\end{abstract}
\newpage
In this note I wish to discuss how the idea that a well 
defined theory underlies all the present QFTs (or QM) can be 
used to deepen our understanding of the quantum theories or 
even to further develop the quantum theories.

Generally, a well-defined physical theory should not contain any 
mysterious parameter and/or structure, i.e., any parameter and/or 
structure appears in the theory should be based on sound physical 
explanations. Once a theory contains some kind of unexplained 
parameters and/or structures, it would go wrong somehow. 

In quantum theory, the quantization procedure itself is not 
well understood. Especially in the case 
of quantization of fields, the procedure itself is over simple
about the short distance structures--just a Delta function to
signal quantum effects. It is known that one would encounter
in the later use of QFT the difficulty like ill-definedness 
and hence one has to introduce additional mysterious structures
via regularization (Reg) in order to calculate quantum corrections. 
The 'deformed' theory would need another mysterious 
procedure--renormalization (Ren).

It is a natural idea that the true description of the nature
should be free of any kind of ill-definedness and unexplained
artificial structures. Suppose we found this complete formulation
(or the underlying theory). Then it is immediate to see that all 
the present quantum theories are in fact low energy (LE) effective 
theories of this underlying one. The conventional quantum theories 
have been obtained in such a way that we first try to establish 
the Hamiltonians or Lagrangians of the relevant phenomena under 
consideration and then take the 'elementary' modes solved from the 
Hamiltonians into the theory via the quantization procedure to 
account for quantum fluctuations. However, the Hamiltonians we 
established {\it must have been wrong} in the UV ends (and/or IR 
ends) of the spectra since these Hamiltonians are just the LE 
limits of certain sections in the underlying theory. The known 
quantization procedure just {\bf uses} these modes that are only 
effective or correct away from the UV (and perhaps IR) ends. In 
other words, the {\sl 'correct' quantization procedure should 
intrinsically take into account the existence of the underlying 
structures rather than using the simplified spectra (wrong at least
at the UV ends)}. 

Now it is clear from the underlying theory point of view that 
we have used structures with something wrong to formulate theories. 
One may remark that since no one knows what the underlying theory 
looks like, what is the point of thinking in terms of the underlying
theory?

First, from the above discussions, we can see that a Reg
procedure is just an artificial substitute for the underlying 
UV structures, thus the results can not be guaranteed to describe 
the phenomena correctly. Since the Ren procedures are again 
artificial operations, the 'renormalized' theories can not be 
assured to describe relevant physics or can not just be the true 
LE limits either. In the conventional formulations, one is seldom 
aware of this danger. The problem is especially severe in the 
unrenormalizable models and more worse in the nonperturbative 
contexts \cite {QMDR}. One might think, in the conventional point 
of view, that once a finite theory is found it is the final one.
However, in the underlying theory point of view, as long as one 
employed some structures or parameters without natural physical 
ground or without confidence that the UV (and/or IR) structure is 
correctly established, then again this 
formulation must have contained certain {\it ad hoc} or 
{\it artificial} ingredients. So, even finite, such formulations 
could not be simply correct. To the author's knowledge, this 
situation has never been pointed out.

Second, the Hamiltonians we have are reasonably correct or healthy
in the effective ranges though they are wrong in the far UV (and/or 
IR) ends. Thus {\it we can make better use of this fact to 
calculate the quantum corrections from the quantities that are in 
fact insensitive to the UV end structures}. This possibility is 
demonstrated by the author in Ref. \cite {YYY}. 
The results are finite (but ambiguous), no specific Reg and Ren
procedures are allowed and no counter terms and bare quantities 
should appear at all. The ambiguities that automatically arise 
just warn us that we are working in the effective theories and 
certain important information about the UV physics is unavailable 
yet. In other words, {\sl the existence of the well-defined 
underlying theory does yield quite powerful consequences for our 
understanding of the quantum theories, especially in getting 
rid of the ridiculous Reg and Ren operations and  
directing our attention to the resolution of the ambiguities}. 
Such kind of efficient use of the standard point of view, as far 
as the author knows, never appeared in the literature.

Third, it is easy to see that the 'elementary' modes in an LE 
effective theory could not continue to be active and 'elementary' 
modes in the higher or lower energy ranges. They would break up 
into (or give way to) new 'elementary' modes at the higher energy 
levels or be somehow confined into new 'elementary' modes (or just 
'sleep') at the lower energy levels as the dynamics is 
different (hence the effective theory is different). Thus the 
propagators and vertices at the UV (or IR) end, if calculated in 
the underlying theory, must have been quite complicated ones due 
to these {\sl dynamics transmutation mechanisms} so that they 
render the result of any loop integration or intermediate state 
summation finite. That is, {\sl the seemingly infinite UV
contributions are in fact 'suppressed by physical mechanisms' or 
should at most contribute a finite part (in the polynomial part
\cite {YYY})}. In a sense, when we
talk about the 'integrating out' of the high energy (HE) 
modes, the 'complicated physical suppression' is intrinsically 
involved and it is imaginable that the
underlying descriptions of the LE effective modes, which are complete 
and well defined, would be very sophisticated ones in terms of 
the underlying parameters. Thus, the conventionally so-called 
'integrating out' of UV or HE modes
necessarily incorporates the complicated but important dynamics
transmutation mechanisms, the conventional formulation using
the HE modes given by the simplified effective models are hence 
problematic. Regrettably, many discussions are still using the 
'wrong' spectrum in this issue and hence ad hoc Reg and/or Ren are 
inevitable and their conclusions need reexamination in principle.

Our technical formulation in Ref. \cite {YYY} made use of the 
following observations for the Feynman amplitudes (FA):
I. There {\it are} convergent graphs or FAs because more 
propagators within a loop lead to better convergence (or are
less sensitive to the UV part of the spectrum);
II. The differentiation with respect to (w.r.t.) the external 
parameters can lead to more internal lines.
These observations can be generalized to other quantum
formulations as long as the summation w.r.t. the intermediate 
states are intrinsically involved. The key point is that
{\it the external (and hence phenomenological) parameters go with
the internal arguments to be integrated or summed over in any 
physically interested amplitude and hence the differentiations 
w.r.t. these external parameters reduce the potential divergence 
of the internal integrals or the summations over intermediate 
states (reduce the contributions from the UV modes) with the help
of the underlying parameters}. In other words, we are
{\it making use of the well-defined sectors of the LE effective 
formulations} and {\it the postulate that a complete theory 
underlies the effective ones} to explore the finite but ambiguous
description in the effective models.
As long as one adopts the underlying theory point of view,
one could find more efficient techniques based
on this physical principle for more general cases.

Now, I wish to discuss some general consequences following
from the underlying theory point of view. The underlying theory, 
if found, must be characterized by the fundamental constants 
(or parameters), $\{\sigma\}$, which are unknown to us yet.
Since the 'elementary' modes in our present models are 
effective ones, all the operators and the state vectors 
or other objects should also depend on the 
$\{\sigma\}$ somehow. 

First let us see what the commutators look like with $\{\sigma\}$
appended. Let us consider any two operators describing the LE
physics which should be well defined in the underlying theory,
say ${\widehat  A}_{\{\sigma\}}$ and ${\widehat  B}_{\{\sigma\}}$, then their
commutator should also be well-defined,
\begin{equation}
[{\widehat A}_{\{\sigma\}} (x;\{\cdots\}),
{\widehat B}_{\{\sigma\}} (y;\{\cdots\})]=
{\widehat C}_{\{\sigma\}} (x,y;\{\cdots\})
\end{equation}
where the space-time arguments for the LE physics are explicitly
labeled and the dots denote the LE phenomenological constants
or parameters like masses, charges, or couplings and so on.

If the two operators are just an LE 'elementary' field and its
conjugate, then the commutator should be the Dirac bracket given
by the underlying theory,
\begin{equation}
[{\widehat {\phi}}_{\{\sigma\}}(x;\{\cdots\}), 
{\widehat {\pi}}_{\{\sigma\}}(y;\{\cdots\})]
=i const. \delta_{\{\sigma\}}(x-y;\{\cdots\}).
\end{equation}
Here we see that the 'elementary' commutator determining the 
canonical quantization procedure should be a rather complicated
function of the space-time separation (we assume as usual that the 
whole nature is still translationally invariant) as the short
distance structures are characterized by the ${\{\sigma\}}$.
One might expect that if we take the LE limit operation on this
commutator we would return to the original commutator with the
right hand of Eq.(2) being the usual delta function, or even we
can get the original form of the operators and their commutators
 if we apply the LE limit operation on the Eq.(1). Generally, 
this is not true as in the effective formulations the simplified
operators (from the LE limit operation) are often singular
or ill defined, that is why people have to resort to Reg in 
practical applications. 

Even this is true for the Dirac bracket, it is still illegitimate
to use it plainly in the construction of the FAs since the  
Dirac bracket just {\it takes the wrong UV structures into the 
quantum theories} as is already pointed out above, which will
inevitably lead to UV divergences. Thus the Dirac bracket 
should be 'corrected' by the underlying UV structures in order
to calculate the quantum corrections in a legitimate way
\cite {YYY}. 

So, if we want to calculate the commutators of more general or
composite operators, we should start with the underlying
theory descriptions characterized by ${\{\sigma\}}$. Before the
calculation is done, one could not apply the LE limit operation
first. Since we do not know the exact formulation of the 
underlying structures, we have to make use of the LE models which
are (almost) correct away from the UV (and/or IR) ends such that
we can get finite but ambiguous expressions. It is natural to
expect that the resulting expressions will be in terms of the LE
phenomenological operators and parameters or constants after the 
LE limit is taken. Again the ambiguities should be fixed
according to physical properties and experiments as stressed 
before. The general technical regime for performing various
calculations of the (field) operators ('elementary' or composite)
needs further construction. 

The most important consequences of the underlying theory postulate
for the effective models are their implications for the novel
properties such as causality, unitarity and locality among others.
All these properties are described in the present quantum theories
by the restrictions of the operators' dependence upon the 
space-time variables. However, it is quite natural to see that in
the underlying theory as a complete theory for at least the UV
limit or short distance limit physics the space-time variables
are again some kind of 'LE phenomenological' quantities, there 
must be deeper structures behind the space-time phenomenon. This
is not a 'quantization claim' of the gravity. (Quantum gravity,
if considered from the underlying theory point of view, is still
an LE effective theory at least due to its bad UV behavior.) 
But we can be sure that the underlying theory will give us a
well-defined and correct description of the gravity which is
automatically a quantum one. 

The physical structures underlying the space-time would imply
that the causality, uniarity and locality, etc. are quite 
profound properties for the LE effective world. In the situation
where the space-time concept tends to breakdown, the causality and
other properties either should be described by other mechanisms
or might cease to be correct somehow. As no one knows the deeper
structures, we stop to make further speculations. But this
conceptual exploration implies that we should reexamine the
causality, unitarity and locality of the effective theories 
with the consideration of the underlying structures to see if
they are consistent with each other or consistent up to what
precision. 

Causality requires that any physical operators should commute 
if they are separated by space-like distance, i.e., the operator 
$C_{\{\sigma\}} (x,y;\cdots)$ in Eq.(1) should vanish if $(x-y)^2$ 
is space-like. It is quite difficult to imagine the breakdown of 
this principle that is so general and reasonable. The author
could only speculate that causality might be a fundamental 
property (or mechanism?) even true when the conventional 
space-time ceases to be effective. Thus, for most energy ranges, 
we might assume the causality for the theories. 
Conversely, we could also 
investigate the general implications of the causality for the 
possible underlying structures.

But the unitarity of an effective theory might not be simply 
assumed as the UV ill-definedness of the effective theories 
often invalidate the naive arguments. The singular potential 
problem provides us such a nontrivial example \cite {QM}. This 
means that we should start with the
existence of the underlying structures to construct a unitary 
effective description. The unitarity issue is also related with 
the Hilbert space structures that will be discussed shortly.
 
The locality is known to hold for at least all the present known
physical quantum theories. But, the underlying structures would 
at least implies that the present formulation of locality is not
correct, especially in the short distance limit (UV) where 
the conventional space-time structure ceases to be true
anymore, or it might breakdown no later than the breakdown of
the conventional space-time.

It seems to be a cheap talk as the underlying 
structures are unavailable. However, the author feels it is 
worthwhile to point out such a scenario which
upon further study might yield quite nontrivial results. It is 
again the underlying theory postulate that leads us to the 
possibility of quite nontrivial structures behind the 
'effective' space-time phenomenon (including the usual quantum 
gravity). 

Next let us discuss the Hilbert space associated with the 
underlying description. The conventional Hilbert spaces for 
various problems should now take into account the influences of 
the underlying structures if one adopts the underlying theory 
approach. Then the Hilbert spaces for describing various LE 
phenomena should be taken as at least various subspaces of a 
whole physical state space and these subspaces, if derived from 
the underlying theory, should also be parametrized somehow by 
the underlying constants and automatically support the LE 
Hamiltonians and other dynamic operators without any 
singularity or ambiguity. In other words, with such Hilbert 
spaces, everything in the effective models should 
be free of any ill-definedness. 

To make physical LE predictions, the vectors in these subspaces 
should also satisfy many conventional conditions as well. But the 
completeness of each of these subspaces should now be defined in 
the underlying theory background, especially in taking into 
account the fact that one dynamics may be related to another 
through dynamic 'phase transition' like mechanisms so that an 
LE model should in principle naturally transmute into the other 
one. As we noted above, different
effective models may use quite different degrees of freedom to
describe physical phenomena, this fact reminds us of further
difficulty in 'unifying' dynamic descriptions. 
In a rough sense, the QCD description of color confinement (if we 
finally find it) must differ from that in terms of the hadrons.

All the LE models might have been ill defined somehow or in some
respects as we did lose the indispensable information about the 
short distance structures in our present formulations. This is in 
contrast to the conventional point of view which says that the LE 
theories should be independent of the short distance structures 
at all. Such a point of view should be corrected as that {\it the 
underlying constants should not appear in the LE formulations but 
they might still influence the LE models through the constants 
arising from the LE limit }\cite {YYY}. Only in very special cases 
can one obtain the LE models in such a way that no ambiguity (no 
constant) appear in the LE limit. However, such models must be of 
quite limited predicting power and the phenomenological constants 
like mass, charge and couplings constants must have been restricted 
by very stringent conditions. A well-known example is the quantum 
mechanics of the Hydrogen atom. In this theory, if the electric 
charge number of the Coulomb potential is permitted to be larger, 
then the Schr\"odinger equation will become ill defined--the 
singular potential problem \cite {Case,QM}. The modified well 
defined theory will depend on an additional parameter specifying 
the influence of the underlying structures on the LE physics 
\cite {QM,Case,YYY}. In fact, the unitarity of such a theory is
established only after the correct Hilbert space for the dynamics
is specified \cite {QM,Case}. Hence we can see the importance of 
the underlying theory in defining a unitary LE theory.

From the underlying theory approach, it is immediate to see that
the conventional spectral representation in quantum theories
should be reformulated, at least we should append the underlying 
constants to the existent formalisms to indicate that the UV ends
of the spectra should be parametrized by these constants. Then 
we can expect that ambiguities instead of UV divergences would 
appear as the underlying descriptions are unavailable yet.

Now, we can see that the existence of the underlying theory does
provide us many important pictures and guidance for the better
understanding of the quantum theories and the problems associated 
with the conventional formulations of the quantum theories.

One might be wondering if there is any relation between this 
underlying theory approach and the recent reformulation of quantum
mechanics in terms of the alternative histories and the coarse 
graining and decoherence of the histories \cite {Gell}.
It seems to me that the new formulation of QM should somehow
automatically take the underlying structures into account, at 
least there should be such possibility. I am not clear how to 
'unify' the two approaches as I am a stranger to this new 
formulation of QM. But both the coarse graining and the fine 
graining of the histories should inherently be based on the 
mechanisms of possible dynamics transmutations. In the 
histories coarse graining approach, the relations 
between the Hilbert spaces corresponding to various grainings 
might be closely related to our discussions just made above. 

One could also consider the interaction between the underlying
theory postulate and the concept of naturalness both in the 
Dirac's sense \cite {Dirac} and in the 't Hooft's sense 
\cite {tHooft}. The underlying theory should yield well-defined
LE descriptions which in turn might exhibit certain 'naturalness'.
However, the two known naturalness criteria should not be used as 
stringent requirements for the effective theories but as useful 
and even important guidance. After all, we are trying to
reveal how the nature evolves but not to let the nature satisfy
our formulation. It is an important as well as interesting task 
to check if the two naturalness criteria (and perhaps new ones) 
are consistent with the existence of the underlying theory in as 
general as possible sense. Again the problem leads us to general 
formulation of the quantum theories automatically taking 
underlying structures into account as mentioned above.

Through our discussions and speculative remarks made above, we do 
not refer to anything in the string theories. Before the final and 
unique formulation (well defined in every aspect) of the string 
theory is firmly established, one might agree that the present 
formulations of the string theories are still 'effective' 
theories (which might be not well defined somehow) of an
underlying theory. Since the quantization procedure in the string
theories is still not new, one would wonder how could such a 
theory be final and truly fundamental. The recent progresses 
about the 'M' or 'F' theory underlies the string theories and 
supergravity \cite {string} could somewhat support such a 
suspicion. But the author does not doubt the importance of the 
investigations in the string theories except that the author 
prefers the belief that {\bf as long as we use the somehow 
simplified mode expansions in the quantization procedure without 
better physical rationale then the theory thus established is 
still an effective one}. In other words, the string theories and its
characterizing parameters might probably be a partial solution of the 
underlying structures but there is little chance for them to
exhaust the underlying structures.

The underlying theory should in principle contain all the 
nontrivial UV and IR structural information that 
each effective theory misses. Then an interesting scenario dawns 
upon us: for each effective model dominating certain energy 
range (say, theory $I_{mid}$), there should exist two other 
effective models (or sectors) that are most adjacent to this 
model from the IR end and UV end respectively (say, $I_{IR}$ 
and $I_{UV}$). It is imaginable that the phenomenological 
parameters in $I_{IR}$ and/or $I_{UV}$ would quite 
nontrivially improve the status of the IR and/or UV behaviors of 
the theory $I_{mid}$. While on the other hand, the $I_{mid}$ 
contains what $I_{IR}$ (or. $I_{UV}$) needs to improve its 
UV (or. IR) behaviors. Put it another way, the active and 
'elementary' modes or fields in $I_{IR}$ will break up in 
$I_{mid}$ and give way to the new 'elementary' modes active 
in $I_{mid}$. Similarly, the 'elementary' modes in $I_{mid}$ 
will go 'hibernating' as the energy goes down while 'new' 
elementary modes 'emerge' to dominate spectra in $I_{IR}$. 
The relation between the elementary modes in $I_{mid}$ and
$I_{UV}$ is in principle just like that between those in $I_{IR}$
and $I_{mid}$. Of course, there may be modes active in several 
successive effective models, some may even be active and stable 
through all energy levels---the 'fossil' modes or fields we 
mentioned in the introduction.

Thus, in a sense, both the IR modes and the UV modes and 
hence the associated phenomenological constants 
characterizing them 'underlie' a QFT (or more generally, a 
quantum theory) if this QFT is ill-defined in the IR and UV 
ends. Thus it is a somewhat different point of view comparing 
with that of the typical reductionism that holds that higher energy 
dynamics is more fundamental than LE ones. Since the IR modes, 
missing from an effective model at a relatively higher energy
level, do {\it underlie} the effective descriptions, the 
underlying structures (both IR and UV ones) and the effective 
structures are in fact unified in an 'organic' way, they depend 
upon each other and they contribute to each other. In principle,
no description at an energy level is more fundamental than the 
others and this is somewhat a {\bf bootstrap} like relation among 
the theories at different levels. Thus if we consider the full 
spectra of the complete description, different modes dominate 
different energy ranges but the complete descriptions of the with 
a certain range needs the 'help' of the underlying structures.

I wish to make some speculative remarks on the distribution 
theory as it is closely related with quantum theories. That the 
distribution theory works necessarily with test function space 
or appropriate measure, if viewed from physical point of view, 
is equivalent to that we need more 'fundamental or underlying 
structures' in order for some singular functions to make sense, 
i.e., a necessity of introducing underlying theory or its 
artificial substitute--regularization. The constructive field 
theory approach, in this sense, also works with a regularization 
effected through the differential properties($C^{k}$) of the 
test functions. From such a hindsight, I wonder if we could
integrate the underlying structure postulate into the mathematical
theory of generalized functions which are almost only meaningful
in physical problems. It will be more amusing to know the relation
between the underlying structure postulate and the nonstandard
analysis in the future.

All the above mentioned discussions, though quite speculative and
not fruitful, are quite important as they concern both our
understanding of the quantum theories and alternatives for 
resolving some difficult yet fundamental problems in the quantum 
theories. Many of these topics are the subjects of our future
investigations both from the conceptual respects and in the 
technical frameworks. One might also derive more implications of 
the underlying theory postulate if she/he could use the existence 
of the underlying theory more efficiently.

\end{document}